\documentclass[10pt]{iopart}
\usepackage{iopams}
\usepackage{graphicx}
\usepackage{color}
\usepackage{amssymb}
\usepackage{cite}
\usepackage{mathptmx}

\begin{document}

\title{Time-resolved qubit readout via nonlinear Josephson inductance}

\author{Georg M. Reuther}
\address{Institut f\"ur Physik, Universit\"at Augsburg,
  Universit\"atsstra{\ss}e~1, D-86135 Augsburg, Germany}

\author{David Zueco}
\address{Instituto de Ciencia de Materiales de Arag{\'o}n y
  Departamento de F\'{\i}sica de la Materia Condensada,
  CSIC-Universidad de Zaragoza, E-50009 Zaragoza, Spain}
\address{Fundaci\'on ARAID, Paseo Mar\'{\i}a Agust\'{\i}n 36,
  50004 Zaragoza, Spain}

\author{Peter H\"anggi}
\address{Institut f\"ur Physik, Universit\"at Augsburg,
  Universit\"atsstra{\ss}e~1, D-86135 Augsburg, Germany}

\author{Sigmund Kohler}
\address{Instituto de Ciencia de Materiales de Madrid, CSIC,
  Cantoblanco, E-28049 Madrid, Spain}

\date{\today}

\begin{abstract}
We propose a generalisation of dispersive qubit readout which provides
the time evolution of a flux qubit observable. Our proposal relies on
the non-linear coupling of the qubit to a harmonic oscillator with
high frequency, representing a dc-SQUID. Information about the qubit
dynamics is obtained by recording the oscillator response to resonant
driving and subsequent lock-in detection. The measurement process is
simulated for the example of coherent qubit oscillations. This
corroborates the underlying measurement relation and also reveals that
the measurement scheme possesses low backaction and high fidelity.
\end{abstract}



\maketitle

\section{Introduction}
\label{sec:intro}

The question of how to gain information about the state of a quantum
system has intrigued researchers since the early days of quantum
mechanics.  With the advent of quantum computation, this fundamental
question became also of practical interest, mainly because the final
stage of a quantum algorithm necessarily is qubit readout. This
task only requires distinguishing between two particular qubit states
and, thus, can be achieved by projective measurements. Nevertheless,
going beyond readout is of interest as well, since one also desires
direct experimental evidence for coherent superpositions
emerging, e.g., from tunnelling oscillations.

In order to obtain a quantum mechanical description of a measurement
process, one usually models the measurement apparatus as a macroscopic
quantum environment, i.\,e., as a heat bath, where the pointer of the
apparatus corresponds to an effective bath coordinate. When
interacting with the central quantum system, the bath acquires
information about the system state. Owing to the macroscopic
nature of the bath, one may assume that already a fraction of the bath
possesses the full information about the effective pointer
coordinate~\cite{Zurek2003a}. Therefore one can obtain knowledge of the
pointer position without violating fundamental laws of quantum
mechanics.

Recently, superconducting quantum circuits have provided a new arena
to test fundamental questions of quantum mechanics in the
laboratory. Prominent examples are the 
demonstration of coherent time evolution in charge
qubits~\cite{Nakamura1999a} and of Berry phases~\cite{Leek2007a},
as well as testing Bell inequalities~\cite{Ansmann2009a}. Above
all, different protocols for quantum measurement were successfully
implemented in circuit quantum electrodynamics~\cite{Sillanpaa2005a,%
Lupascu2006a,Lupascu2007a, Schuster2007a}.
For a superconducting solid-state qubit, the practical measurement of
one of its coordinates is performed by coupling it to a macroscopic
environment, as given by external circuitry, via a quantum point
contact~\cite{Ashhab2009a,Ashhab2009b} or a harmonic
oscillator. Depending on the setup, the oscillator is realised by a dc
superconducting quantum interference device
(SQUID)~\cite{Chiorescu2004a} or a superconducting 
resonator~\cite{Wallraff2004a}. In both cases, the resonance
frequency of the oscillator depends on the qubit state. Consequently,
the response of the oscillator to a close-to-resonant ac-excitation
possesses a phase shift which can be measured, and from which one can
infer the qubit state. First experiments in this direction worked with
an oscillator whose frequency was much lower than the qubit
splitting~\cite{Grajcar2004a,Sillanpaa2005a,Johansson2006a}. More 
recent experiments~\cite{Lupascu2006a,Filipp2009a} operated in the
so-called dispersive regime, where the oscillator frequency and the
qubit splitting are of the same order, while their detuning is still
larger than their mutual coupling strength. A crucial detail is that the
oscillator frequency and bandwidth naturally limit the
time resolution in such qubit measurements. Thus, using the said
schemes with slow oscillators, it is only possible to extract
time-averaged information about the qubit state in general, but there
is no possibility to resolve its dynamics in time.

Recently, a first step towards a time-resolved measurement of qubit
dynamics has been proposed \cite{Reuther2009a,Reuther2011a}: When a weak
high-frequency field acts directly on the qubit, the reflected signal
acquires a time-dependent phase shift by harmonic mixing. Lock-in
amplification of the reflected signal then allows obtaining
information about the qubit dynamics.
In this work, we combine both approaches and extend the scheme of
Refs.~\cite{Reuther2009a, Reuther2011a} to a qubit coupled to a driven
high-frequency oscillator. A measurement protocol for such a setup is
particularly appealing because an oscillation mode is part of most
recent superconducting qubit designs. Moreover, the oscillator serves
as filter for quantum noise and, thus, reduces qubit decoherence.
Here, we focus on a flux qubit embraced by a dc-SQUID, whose
fundamental frequency may even be tunable to some
extent~\cite{Bertet2005a,Bertet2005b}. As a particular feature of this
realisation, the qubit-oscillator coupling is non-linear in the
oscillator coordinate, that is, the coupling possesses both a
significant linear and quadratic contribution. It will turn out that for
realistic parameters, the measurement scheme relies on the quadratic
part of the coupling.

The paper is structured as follows. In section~\ref{sec:model}, we
introduce our model and discuss dispersive qubit readout in 
generalised terms. The central relation upon which our measurement
scheme relies is derived in section~\ref{sec:response}.
Section~\ref{sec:numerics} is devoted to numerical studies in which we
test our measurement relation and work out quantitatively measurement
fidelity and backaction. Furthermore, we provide an estimate of the
signal-to-noise ratio and discuss realistic parameters for a possible
experimental implementation. The appendix contains details about the
derivation of the measurement relation, the input-output
formalism~\cite{Gardiner1985b} and the Bloch-Redfield master equation
which we use for obtaining numerical results.

\section{Dissipative qubit-oscillator model}
\label{sec:model}

\subsection{System-bath model}
\label{sec:sysbath}

\begin{figure}[t]
\centering
\includegraphics[scale=.5]{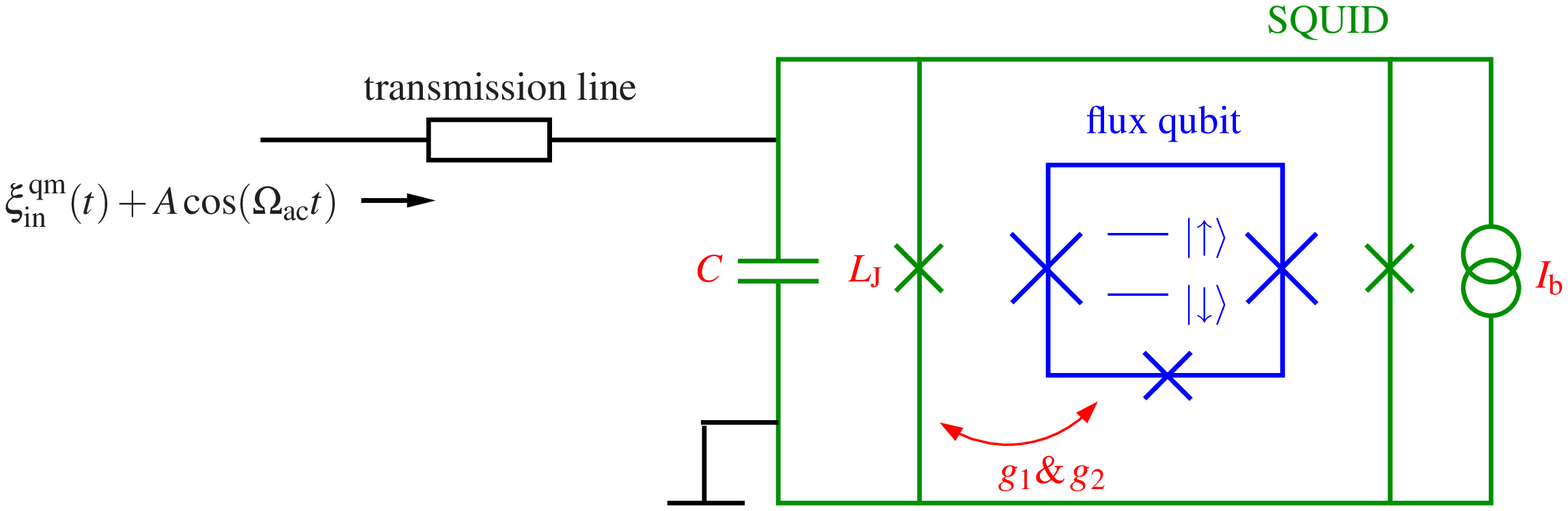}
\caption{
Sketch of the flux qubit (blue) coupled to a dc-SQUID. The interaction is
characterised by the linear coupling $g_1$,
which depends linearly on the SQUID bias current $I_\mathrm{b}$,
and the quadratic coupling 
$g_2$. The SQUID with Josephson inductance $L_\mathrm{J}$ is shunted
by a capacitance $C$. The frequency shift of the resulting harmonic
oscillator (green) can be probed by external resonant ac-excitation $A \cos
(\Omega_{\mathrm{ac}}t)$ via the transmission line (black), in which the
quantum fluctuations $\xi_{\mathrm{in}}^{\mathrm{qm}} (t)$ are also
present.
}\label{fig:setup} 
\end{figure}
%
We consider a superconducting flux qubit coupled to a
SQUID~\cite{Lupascu2007a} as sketched in figure~\ref{fig:setup}. The
SQUID is modelled as a harmonic oscillator, which gives rise to the
Hamiltonian \cite{Bertet2005a,Lupascu2006a,Lupascu2007a,Serban2008a}
\begin{equation}
\label{eq:H_qc}
\hspace{-2em}
\mathcal{H}_0 =  \frac{\hbar \omega_\mathrm{qb}}{2} \sigma_z 
   + \hbar \Omega \Big( a^\dagger a +\frac{1}{2} \Big)
   + \hbar( \sigma_z \cos \theta - \sigma_x \sin \theta )
 \left[ g_1 (a + a^\dagger) + g_2 (a + a^\dagger)^2 \right] .\,
\end{equation}
The first term represents the qubit with energy splitting $\hbar
\omega_\mathrm{qb} = \hbar(\epsilon^2 + \delta ^2)^{1/2}$ and the
mixing angle $\theta = \arctan(\delta/\epsilon)$ which depends on the
controllable qubit bias energy $\epsilon$ and the qubit gap energy
$\delta$, while $\sigma_{x,z}$ denote the Pauli matrices.
The second term describes the oscillator with frequency $\Omega$ and
the bosonic creation and annihilation operators $a^\dagger$ and $a$,
respectively. The qubit couples to the oscillator in two ways. First,
via dipole interaction with strength $g_1$, which is linear
in the oscillator coordinate $a + a^\dagger$.  Up to order $g_1^2$, this
causes a frequency shift for both the oscillator and the qubit.  The second
coupling term proportional to $g_2$, by contrast, is quadratic in the
oscillator coordinate. Its physical origin is a non-linear
Josephson inductance which depends on the magnetic flux, by which the
SQUID is penetrated~\cite{Bertet2005a,Lupascu2007a}. This term
provides a frequency shift already in first order of $g_2$. The
interaction coefficients $g_1$ and $g_2$ can be controlled to some
extent, as an expansion of the qubit-SQUID interaction to second order
in the oscillator coordinate demonstrates~\cite{Bertet2005a}. In
detail, for a small SQUID bias current $I_\mathrm{b}$, the coupling
coefficient $g_1$ depends linearly on $I_\mathrm{b}$, whereas  $g_2$
is independent of the latter, as we discuss in
section~\ref{sec:experimental}. For $I_\mathrm{b}=0$, one even obtains
$g_1=0$, such that the qubit couples only to the square of the
oscillator position~\cite{Bertet2005a,Lupascu2007a}.

Regarding a time-resolved measurement of the qubit dynamics via the
oscillator, it will turn out that for realistic parameters of flux
qubits, this quadratic coupling is crucial, while the linear coupling
turns out to be typically too weak.  For common circuit-QED setups
using charge and flux qubits coupled to a transmission line
resonator~\cite{Blais2004a, Wallraff2004a,
Chiorescu2004a,Mariantoni2008a}, not only $g_1$ but also $g_2$ is too
small. Thus, we henceforth focus on setups of flux qubits coupled to
SQUIDs possessing a sizable quadratic coupling, as described above.

The qubit-SQUID system is further coupled to external circuitry, which
acts as a dissipative environment and is modelled by the system-bath
Hamiltonian~\cite{Leggett1987a,Hanggi1990a,Weiss1999a,Makhlin2001b}
\begin{equation}
  \label{eq:sb}
   \mathcal{H} =  \mathcal{H}_0  + Q \sum_k \hbar c_k
   (b_k^{\vphantom{\dagger}} + b_k^{\dagger}) + \sum_k \hbar \omega_k
   \Big( b_k^{\dagger}b_k^{\vphantom{\dagger}} + \frac{1}{2} \Big)
   \; .
\end{equation}
Here, $Q = a^\dagger+a$ is the oscillator coordinate, such that the
interaction term represents the inductive coupling between the qubit
and the flux degree of freedom of the SQUID.  The system-bath
interaction can be fully characterised by the spectral density
$J(\omega) = \pi \sum_k |c_k|^2 \delta(\omega-\omega_k) $ which is
proportional to the real part of the effective impedance of the
environment~\cite{Ingold1992a}. Here we assume an ohmic spectral
density, $J(\omega) = \alpha \omega$, for which the dimensionless
damping strength $\alpha$ can be interpreted as effective
resistance~\cite{Yurke1984a,Devoret1995a,Makhlin2001a}.

\subsection{Qubit-oscillator interaction in the dispersive limit}
\label{sec:disp}

We are interested in the dispersive limit which is characterised by a
detuning $\Delta = \Omega - \omega_\mathrm{qb}$ larger than the
qubit-oscillator couplings,
\begin{equation}
  g_1,g_2 \ll |\Delta| ,
  \quad \Delta= \Omega - \omega_{\mathrm{qb}} \; .
   \label{eq:dispersive0}
\end{equation}
It is then convenient to go to the dispersive picture via the
unitary transformation \eref{eq:drivenosc-Schrieffer}.
As detailed in~\ref{app:dispersive}, this yields the
effective Hamiltonian~\cite{Klimov2000a,Klimov2003a}
\begin{equation}
\label{eq:Heff10}
\bar{\mathcal H}_0
= \mathcal U^\dagger \mathcal H_0 \mathcal U
= \hbar\bar\Omega 
\Big( \bar a^\dagger \bar a +\frac{1}{2}\Big)
+ \frac{\hbar\omega_{\mathrm{qb}}}{2}\sigma_z \;,
\end{equation}
where the transformed bosonic operators $\bar a$ and $\bar a^\dagger$
are defined in equation \eref{eq:abar}.  The qubit-oscillator coupling
has been removed formally by shifting it to the operator-valued
oscillator frequency
\begin{equation}
  \label{eq:Omegabar0}
  \bar \Omega
 = \Omega\sqrt{1 + \frac{4\bar \omega}{\Omega}} \, ,
\end{equation}
where the overbar denotes the dispersive picture, while the qubit operator
\begin{equation}
  \label{eq:omegabar0}
 \eqalign{
  \bar \omega =&
  \frac{g_1^2}{2}  \sigma_z
  \Big(\frac{1}{\Delta} - \frac{1}{\Omega +
      \omega_{\mathrm{qb}} } \Big) \sin^2 \theta
  + \frac{g_1^2}{2} \sigma_x
    \Big(\frac{1}{\Delta} - \frac{1}{\Omega +
        \omega_\mathrm{qb}}  \Big)  \cos\theta \sin\theta\\ &
 +   g_2 (\sigma_z \cos \theta - \sigma_x \sin\theta ) \,
    ,}
\end{equation}
determines the coupling.
The interpretation of equations~\eref{eq:Omegabar0} and
\eref{eq:omegabar0} is that the oscillator frequency depends on the
qubit state.  This allows dispersive qubit readout by measuring the
associated phase shift of the oscillator response upon resonant
driving, like in the case of a classical oscillator that
  is driven by an external force. In particular, assuming $\cos\theta =
0$ and $g_2=0$, equation~\eref{eq:Omegabar0} predicts the frequency
shift $\bar \Omega = \Omega + \sigma_z g_1^2[1/\Delta - 1/(\Omega +
\omega_\mathrm{qb})]^{-1}$.
The last contribution in $\bar \Omega$ stems
from counter-rotating terms in the qubit-oscillator interaction. These
must be accounted for the case of large detuning $\Delta$ where a
rotating-wave approximation produces inaccurate
results~\cite{Zueco2009b}. Depending on the qubit expectation value
$\langle \sigma_z \rangle$, the oscillator is red or blue
detuned. Thus, we obtain in this limit the well-known qubit-dependent
phase shift corroborated in various experimental
realisations~\cite{Blais2004a,Sillanpaa2005a,Lupascu2006a,%
Lupascu2007a,Schuster2007a}. There, however, the oscillator frequency
was smaller than the qubit splitting, $\Omega \ll \omega_\mathrm{qb}$.
As a consequence, it was only possible to obtain
\textit{time-averaged} information about the qubit state.

Now the goal of this paper is a generalisation of dispersive qubit
readout such that \textit{time-resolved}
information about the qubit state can be obtained as well. This obviously
requires oscillator frequencies and bandwidths larger
than the qubit transition 
frequency, that is, $\Omega \gg \omega_\mathrm{qb}$. We emphasise that
equations~\eref{eq:Omegabar0} and~\eref{eq:omegabar0} are nevertheless
valid as long as the coupling constants are small enough to fulfil
condition~\eref{eq:dispersive0}; for details see \ref{app:dispersive}
and Ref.~\cite{Zueco2009b}. If the qubit dynamics is much slower than
the oscillator, the qubit can be treated within an adiabatic
approximation. This means that the qubit dynamics is assumed to be
constant during one oscillator period. In turn, the time evolution
of the oscillator depends on the instantaneous qubit state.
Then the Schr\"odinger-picture operators
$\sigma_{x,z}$ in equation~\eref{eq:omegabar0} can be replaced
by their time-dependent expectation values, and the operator-valued
quantity $\bar \omega$ is substituted by
\begin{equation}
  \label{eq:omegabar1}
 \eqalign{
    \bar \omega (t) &= \frac{(g_1 \sin \theta)^2}{2} 
    \left(\frac{1}{\Delta} - \frac{1}{\Omega
      +\omega_\mathrm{qb} } \right) \langle \sigma_z \rangle_t 
     +\frac{g_1^2}{2}  \cos \theta \sin \theta 
    \left(\frac{1}{\Delta} - \frac{1}{\Omega
   +\omega_\mathrm{qb} } \right) \langle \sigma_x  \rangle_t \\
    & \quad +  g_2 \big(\cos \theta \langle \sigma_z \rangle_t - \sin
    \theta \langle \sigma_x \rangle_t \big) \; .
 }
\end{equation}
Equation~\eref{eq:omegabar1} implies that information about the
time-dependent qubit state is encoded in the effective oscillator
frequency $\bar \Omega \equiv \bar \Omega (t)$. This gets expressed
as a slow parametric modulation in time, like for a
  parametric oscillator. In detail, the instantaneous qubit state
  enters via the qubit expectation values $\langle \sigma_{x,z}
  \rangle_t \equiv \mathrm{Tr_\mathrm{qb}} \{ \sigma_{x,z} \rho_0 (t)
  \}$, where $\mathrm{Tr_\mathrm{qb}}$ denotes the partial trace over
  the qubit degrees of freedom. The time dependence, indicated by the
  subscript $\langle \ldots \rangle_t$, stems from the evolution of
  the total qubit-oscillator state $\rho_0(t)$ under the effective
  system-bath Hamiltonian~\eref{eq:drivenosc-sbeff}.

As an important intermediate result, the found modulation
of $\bar \Omega$ in time can be traced back to the qubit
dynamics in the absence of the driving. This enables measuring the
qubit's time evolution via the oscillator response to resonant
driving.

\section{Time-resolved measurement of the qubit dynamics}
\label{sec:response}

The qubit-oscillator Hamiltonian in the dispersive picture,
equation~\eref{eq:Heff10}, together with the effective, modulated
frequency \eref{eq:Omegabar0} already indicates that the oscillator
detuning may contain information about the qubit dynamics.  As in the
case of the traditional dispersive readout, we consider the response
of the system to an ac-field that is resonant with the oscillator.
Physically, the situation corresponds to that of a classical
mechanical oscillator driven by an external periodic force. 
Owing to the only weak dissipation, the response is manifest in the
phase of the reflected ac driving.
In the following, we establish a relation between this phase
and a time-dependent qubit expectation value.  This relation will form
the basis of our measurement protocol.

\subsection{Response of the qubit-oscillator compound to resonant driving}
\label{sec:inout}

In the theory of optical cavities, the response to an external ac
excitation is conveniently calculated with the input-output
formalism~\cite{Gardiner1985b, Gardiner2004a}.  This formalism has
also been applied to quantum circuits~\cite{Grajcar2004a,
Sillanpaa2005a, Johansson2006a, Reuther2009a}. Its cornerstone is the
relation
\begin{equation}\label{eq:in-out1}
 \xi_{\mathrm{out}}(t) - \xi_{\mathrm{in}}(t) = 2 \alpha  \dot Q \; ,
\end{equation}
formulated in the Heisenberg picture and derived
in~\ref{app:input-output}.  It relates the incoming and the outgoing
fluctuations of the transmission line, $\xi_\mathrm{in/out}(t)$, to
the time-derivative of the system-bath coupling operator, which in our
case is $Q = a + a^\dagger$. The dimensionless dissipation strength
$\alpha$ of the ohmic spectral density quantifies the coupling between
the oscillator and the electric environment and, thus, appears as
prefactor. An ac-driving corresponds to a coherently excited incoming
mode, such that the fluctuations can be separated into quantum
fluctuations $\xi_\mathrm{in}^\mathrm{qm}(t)$ and a deterministic
component $A \cos (\Omega_\mathrm{ac} t)$. Here, the deterministic
part is an ac-field in resonance with the bare oscillator,
$\Omega_\mathrm{ac} = \Omega$, such that
\begin{equation}\label{eq:in-det}
\xi_{\rm in} (t)
= \xi_{\rm in}^{\rm qm} (t) + A \cos (\Omega t) ,
\end{equation}
which implies the expectation value $\langle\xi_{\rm in}(t)\rangle =  A
\cos(\Omega t)$.  Then the input-output relation~\eref{eq:in-out1}
becomes $\xi_\mathrm{out} (t)  = \xi_\mathrm{in}^\mathrm{qm} (t) +
A\cos(\Omega t)  + 2 \alpha  \dot Q $ and, thus, the corresponding
expectation value of the outgoing signal reads
\begin{equation}
\label{eq:in-out-total1}
\langle \xi_\mathrm{out} (t) \rangle 
= A\cos(\Omega t) + 2 \alpha \langle \dot Q \rangle .
\end{equation}
Also here, it is convenient to work in the dispersive picture obtained
by the unitary transformation \eref{eq:drivenosc-Schrieffer}.  While
this leaves the environment operators unchanged, the coordinate by
which the oscillator couples to the environment changes as
$Q \to \bar Q=\bar a+\bar a^\dagger -(\lambda_\Delta-\lambda_\Sigma)
\sigma_x +2\lambda_\Omega\sigma_z$; see equation~\eref{eq:drivenosc-Qeff}.
The time-derivative $\rmd{\bar{Q}}/\rmd t$ can be obtained from the commutator
of $\bar Q$ with the Hamiltonian~\eref{eq:Heff10} augmented by a term
that describes the driving.  This yields terms of the order $\Omega$
and terms with prefactors $\omega_\mathrm{qb}$ and
$g_1/\Delta$.  For a fast oscillator, the latter
terms can be neglected, and we obtain the equation of motion
\begin{equation}
\frac{\rmd^2}{\rmd t^2}{\bar Q}
+ 2 \alpha \bar \Omega \frac{\rmd}{\rmd t}{\bar Q}
+ \bar \Omega^2 \bar Q
= - 2 \bar \Omega \left[ \xi^\mathrm{qm}_\mathrm{in} (t) + A \cos
  (\Omega t) \right] \; .
\end{equation}
This linear, inhomogeneous equation is readily solved with the help of
the Green's function for the dissipative harmonic oscillator.
Inserting the resulting $\rmd{\bar Q}/\rmd t$ into the input-output
relation~\eref{eq:in-out-total1} and neglecting transient terms
yields for the expectation value of the outgoing signal the expression
\begin{equation}
  \label{eq:xiout1}
   \langle\xi_\mathrm{out} (t)\rangle  
   = A \cos\{\Omega t - \varphi(t)\} \;.
\end{equation}
Owing to the weak dissipation, the system energy is almost preserved,
such that the amplitude of the incoming and the outgoing signal are
practically the same. The phase difference
\begin{equation}
  \label{eq:phi}
  \varphi(t) = \arctan\Bigg(
  \frac{- 4 \alpha \bar \Omega \Omega \left(\bar \Omega^2 - \Omega^2\right)}
       {(\bar\Omega^2-\Omega^2)^2 - 4 \alpha^2\bar\Omega^2 \Omega^2} \Bigg) 
\approx \frac{\bar \Omega^2-\Omega^2}{\alpha\Omega\bar\Omega}
\end{equation}
stems from the coupling to the qubit which detunes the oscillator,
while the slow time evolution of the qubit renders the phase
time-dependent.  The approximation is valid if the qubit-oscillator
couplings are smaller than the oscillator damping rate, i.e., for
$g_1,g_2 \ll \alpha\Omega$. In other words, the first term in the
denominator is negligible, since the qubit-induced frequency
shift $\bar\Omega-\Omega$ is of the order $g_{1,2}$. This also ensures
$\varphi\ll1$ and, thus, $\varphi \approx\tan\varphi$.
Next, we insert the effective frequency~\eref{eq:Omegabar0} together
with equation~\eref{eq:omegabar1} and obtain to second order in $g_1$
and first order in $g_2$ the phase shift
\begin{equation}
  \label{eq:dr-osc-phase}
\eqalign{
     \varphi (t) &= \frac{2 g_1^2}{\alpha \Omega} 
     \left(\frac{1}{\Delta} - \frac{1}{\Omega +
        \omega_{\mathrm{qb}} } \right)
    \Big[  \sin^2 \theta 
     \langle \sigma_z\rangle_t
     +  \cos \theta \sin \theta 
         \langle \sigma_x \rangle_t  \Big ] \\ 
    &\quad + \frac{4 g_2}{\alpha \Omega} 
    \left(\cos \theta \langle \sigma_z \rangle_t 
    - \sin \theta \langle \sigma_x \rangle_t \right) \; .
}
 \end{equation}
This central relation forms the basis for our non-invasive qubit
measurement via a resonantly driven harmonic oscillator. It
identifies a set of qubit observables, which generate the
low-frequency system dynamics, as the cause of a small phase shift
between the ingoing and outgoing signal. In other words,
equation~\eref{eq:dr-osc-phase} enables one to monitor the qubit
dynamics by continuously measuring the phase shift $\varphi(t)$ with
suitable experimental techniques.

By evaluating the prefactor for specific setups, we will see below
that our measurement scheme is particularly feasible for flux qubits.
In this case, the last term of the phase shift~\eref{eq:dr-osc-phase}
dominates, and one measures the qubit variable $\sigma_z\cos\theta -
\sigma_x\sin\theta$, i.e., the flux degree of freedom by which the
qubit couples to the SQUID; cf.\ the model Hamiltonian \eref{eq:H_qc}.

\subsection{Static versus dynamical phase shift}
\label{sec:drivenosc-staticdyn}

The terms entering the phase shift $\varphi(t)$ may be static as well
as dynamical. In the first instance, this depends on whether or not the
related qubit observables undergo any time evolution.  At this point,
further insight is obtained by a closer look to the Heisenberg
equations of motion for the qubit operators $\sigma_x$ and $\sigma_z$.
In the dispersive picture, valid under conditions \eref{eq:dispersive0},
they are derived from the effective Hamiltonian $\bar{\mathcal H}_0$,
given by equation~\eref{eq:Heff10}, and read
\begin{eqnarray}
  \dot \sigma_x 
  &=& \frac{\rmi}{\hbar}[ \bar{\mathcal H}_0, \sigma_x] 
  = - \omega_\mathrm{qb} \sigma_y \; ,
  \\
  \dot \sigma_z
  &=& \frac{\rmi}{\hbar}[ \bar{\mathcal H}_0, \sigma_z] 
  = 0 \; .
\label{eq:dotz}
\end{eqnarray}
Thus, in the dispersive qubit-oscillator coupling limit (see
section~\ref{sec:disp}), the observable $\sigma_z$ is a constant of
motion. As a consequence, those contributions to $\varphi(t)$ that
depend on $\langle\sigma_z\rangle_t \equiv \langle\sigma_z\rangle_{\rm
  const} $ are time-independent. This corresponds again to the
established scheme for non-invasive qubit state readout.

On the contrary, the observable $\sigma_x$ possesses a non-trivial
time dependence generated by $ \bar{\mathcal H}_0$. Thus,
$\langle\sigma_x\rangle_t$ renders the phase shift $\varphi(t)$
dynamical. This, in turn, enables a time-resolved single-run
measurement of the unitary qubit evolution by means of the qubit
observable $\sigma_x$. According to our measurement
relation~\eref{eq:dr-osc-phase}, the dynamical phase signal has the
amplitude
\begin{equation}\label{eq:phaseresolution}
\varphi_\mathrm{max}^x = \frac{2}{\alpha \Omega}
\left| g_1^2 \sin \theta \cos \theta \left(
  \frac{1}{\Delta}  -\frac{1}{\Omega +
    \omega_\mathrm{qb}} \right) - 2 g_2 \sin  \theta \right|
 \; .
\end{equation}
Interestingly, $\varphi^x_\mathrm{max}$ is reciprocal to the damping
strength and the oscillator half-bandwidth $\alpha\Omega$.
Thus, a large oscillator frequency together with strong damping lead
to reduced angular visibility.  On the other hand, the adiabatic
treatment of the qubit underlying relation~\eref{eq:dr-osc-phase} becomes
invalid if either $\Omega$ or $\alpha$ are too small. Moreover, the
input-output relation~\eref{eq:in-out-total1} crucially relies on
finite damping. Thus, appropriate choices for $\Omega$ and $\alpha$
need to be based upon a compromise between good phase resolution and
the validity of our approximations. We go into further detail about
this issue when discussing the measurement quality in
section~\ref{sec:drivenosc-fid}.

It is important to note that the amplitude $\varphi_\mathrm{max}^x$
possesses contributions from both the linear and the quadratic
qubit-oscillator interaction of Hamiltonian~\eref{eq:H_qc}. The first
term on the r.\,h.\,s.\ of equation~\eref{eq:phaseresolution} stems 
from the linear interaction  characterised by the coupling coefficient
$g_1$. Like the effective Hamiltonian~\eref{eq:drivenosc-Heff1},
this contribution is of second
order in the dispersive parameter $g_1/\Delta$. Due to the minus sign
inside the round brackets, it is additionally minimised, given that
$\Delta / (\Omega +  \omega_\mathrm{qb}) \approx 1$ for  large
detuning $\Delta$.  Thus, for a Cooper-pair box or a flux qubit 
coupled to a high-frequency transmission line resonator, where
the qubit-oscillator coupling merely depends on the circuit
characteristics and therefore is purely linear, i.\,e.\ $g_2 =
0$, the maximum amplitude $\varphi_\mathrm{max} ^x$ drops below
any useful level.

On the contrary, a finite quadratic qubit-oscillator interaction
$g_2 > 0 $ ensures a noticeable phase signal, independent
of the detuning $\Delta$, and even if $g_2 \ll g_1$. 
  Formally, this feature arises from 
  Hamiltonian~\eref{eq:H_qc}, where the quadratic coupling term
  already generates a frequency shift in zeroth order perturbation
  theory. If the
qubit-oscillator interaction is transverse, that is, $\epsilon = 0$,
the phase resolution is maximised, whereas it vanishes for purely
longitudinal coupling, i.e., for $\delta =0$. Hence the presence of a
non-linear qubit-oscillator interaction, as provided by a non-linear
SQUID Josephson inductance, turns out to be a crucial ingredient.

\section{Measurement quality}
\label{sec:numerics}

Still it remains to corroborate the central measurement
relations~\eref{eq:dr-osc-phase} and \eref{eq:phaseresolution},
respectively, via comparing them with the phase shift obtained by
simulating the actual measurement process. In doing so, we restrict
ourselves to the fundamental example of coherent qubit oscillations.
For the numerical treatment of the qubit-oscillator state, we employ
the quantum master equation~\eref{eq:blochredfield} derived from the
full dissipative qubit-oscillator-bath Hamiltonian~\eref{eq:sb}.  For
a realistic evaluation, we use parameters similar to those of the
experiment reported in references~\cite{Bertet2005a,Bertet2005b}.
Furthermore, we employ $N=10$ oscillator states, which turns out to be
sufficient to reach numerical convergence.

\subsection{Time-resolved measurement of unitary qubit evolution}
\label{sec:qubit-evolution}
\begin{figure}[tb!]
  \centering
  \includegraphics[scale=1.]{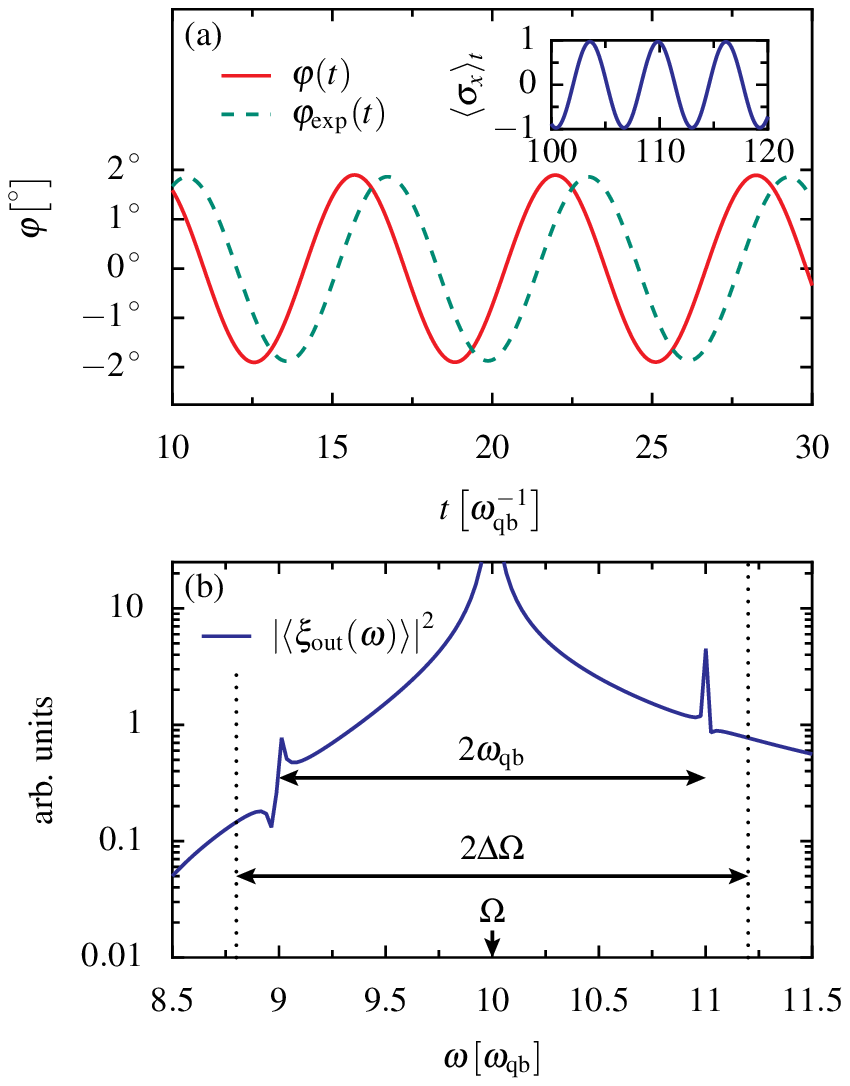}
  \caption
    { Time-resolved measurement of coherent qubit
      oscillations at the degeneracy point $\epsilon = 0$. The full
      qubit-oscillator state was simulated with the quantum master
      equation~\eref{eq:blochredfield} with $N=10$ oscillator states
      and the parameters $\Omega = \Omega_\mathrm{ac} = 10\,
     \omega_\mathrm{qb}$, $g_1 = 0.1\, \omega_\mathrm{qb}$, $g_2 = 0.01\,
     \omega_\mathrm{qb}$, $A = 1.0\, \omega_\mathrm{qb}$. 
      The dimensionless oscillator dissipation strength
      is $\alpha = 0.12$. The resonator bandwidth is given by $2\alpha
      \Omega = 2.4\omega_\mathrm{qb}$.
 (a) Lock-in amplified phase $\varphi_\mathrm{exp} (t)$ (dashed
     green lines), compared to the estimated phase $\varphi (t)$ (solid red
     line) of the outgoing signal $\langle \xi_\mathrm{out} (t) \rangle$.
     Here, $\varphi (t) \propto \langle \sigma_x \rangle_t$ [cf.\
     equation~\eref{eq:dr-osc-phase}], which is corroborated by the inset
     showing that $\langle \sigma_x \rangle_t$ performs oscillations
     with (angular) frequency $\omega_\mathrm{qb}$.
     (b) Power
     spectrum $ \langle\xi_\mathrm{out} (\omega) \rangle$ for the
     resonantly driven oscillator (blue solid line).  The sidebands
     stemming from the qubit dynamics are visible at frequencies
     $\Omega \pm \omega_\mathrm{qb}$. In order to extract the 
     phase information, we apply a Gaussian window function with respect to
     the frequency window of half-width $\Delta \Omega
     = 1.2\omega_\mathrm{qb}$, which turns out to be the
       optimal value for the measurement bandwidth.
}
  \label{fig:qubit-osc-phase-spectrum}
\end{figure}

If the qubit is only weakly coupled to the oscillator, and if the
latter is driven only weakly, the qubit's time evolution is
rather coherent (see section~\ref{sec:sn} on qubit decoherence).
For this 
scenario, figure~\ref{fig:qubit-osc-phase-spectrum}(a) depicts the
time-dependent phase $\varphi(t)$ computed with the measurement
relation \eref{eq:dr-osc-phase}, while the inset confirms its
proportionality to the qubit expectation value $\langle \sigma_x
\rangle_t$. For a comparison, we wish to recover this phase
information directly by analysing the outgoing signal $\langle
\xi_\mathrm{out}(t) \rangle$, as given by equation
\eref{eq:in-out-total1}.  In an experiment, this can be
achieved by lock-in techniques which we mimic in the following
way~\cite{Scofield1994a}: First, we focus on the associated spectrum
$\langle \xi_\mathrm{out} (\omega)\rangle$ depicted in
figure~\ref{fig:qubit-osc-phase-spectrum}(b). It reflects the qubit
dynamics in terms of two sidebands around the central peak
related to the oscillator frequency, here chosen as $\Omega =
10\,\omega_\mathrm{qb}$. The dissipative influence of the
environment, modelled by a transmission line (see figure
\ref{fig:setup}), is reflected in a broadening of this peak. The
corresponding oscillator bandwidth is given as $2\alpha \Omega$, where
$\alpha$ denotes the dimensionless damping strength; see
\ref{app:QME}. Here, we recall that the oscillator is driven
resonantly by the external driving signal $A \cos(\Omega_\mathrm{ac}
t)$, that is, $\Omega = \Omega_\mathrm{ac}$.  In the time domain, the
sidebands correspond
to the phase-shifted signal $\langle\xi_\mathrm{out}(t)\rangle = A
\cos(\Omega t -\varphi_\mathrm{exp}(t))$ with slowly time-dependent
phase $\varphi_\mathrm{exp}(t)$.  In order to obtain this phase
$\varphi_\mathrm{exp}(t)$, we select the spectral data from a
frequency window of size $2\Delta\Omega$ centred at the oscillator
frequency $\Omega$, which means that $\langle \xi_\mathrm{out}
(\omega)\rangle$ is multiplied with a Gaussian window function
$\exp(-(\omega-\Omega)^2/\Delta \Omega^2)$.  We choose for the window size
the resonator bandwidth, $\Delta\Omega = \alpha\Omega$, which turns
out to suppress disturbing contributions from the low-frequency qubit
dynamics.  Finally, we centre the clipped spectrum at zero frequency
and perform an inverse Fourier transform to the time domain.
If the phase shift $\phi_\mathrm{ext}$ was constant,
one could use a much smaller measurement bandwidth.  Then the outcome
of the measurement procedure would correspond to homodyne detection
\cite{Gardiner2004a} of a quadrature defined by the phase shift and
yield a value $\propto\cos\phi_\mathrm{exp}$.

Figure~\ref{fig:qubit-osc-phase-spectrum}(a) reveals the good
agreement of the resulting
$\varphi_\mathrm{exp}(t)$ with the prediction of our
measurement relation, $\varphi(t) \propto \langle\sigma_x\rangle_t$,
at angular resolutions of $1$--$2^\circ$. Good agreement is already
obtained for a oscillator frequency $\Omega = 10 \,\omega_\mathrm{qb}$,
which obviously represents a good compromise between the validity of
the adiabatic approximation (see section~\ref{sec:disp}) and a
sufficiently strong signal.  There is even some room for obtaining a
stronger phase signal since the dissipation strength $\alpha$ still
can be reduced without violating the validity range of our theory
as long as $\omega_\mathrm{qb} \lesssim \Delta \Omega$.

Setting either $g_1$ or $g_2$ to zero (not shown) reveals that the
non-linear coupling $g_2$ is responsible for the good agreement of
the phase shifts in figure~\ref{fig:qubit-osc-phase-spectrum}(a).
Thus, the whole protocol is mainly applicable to flux qubits coupled
to SQUIDs. For charge qubits, by contrast, the typical values for
$g_2$ are too small.  Furthermore, we have verified that the visible
constant delay between both phases $\varphi (t)$ and
$\varphi_\mathrm{exp} (t)$ does not depend on the selected parameters,
while its detailed origin remains unexplained.

\subsection{Measurement characterisation: fidelity and backaction}
\label{sec:drivenosc-fid}

The validity of relation~\eref{eq:dr-osc-phase} for the phase
$\varphi (t)$ is naturally limited to specific parameter ranges due to
the various underlying approximations made. The main crucial assumptions 
to justify the adiabatic treatment of the qubit are a large
qubit-oscillator detuning $\Delta/\Omega \simeq 1$ and
weak mutual interaction, $g_2 \ll g_1 \ll \Delta$. Furthermore, the
oscillator damping $\alpha$ is assumed to stay within the limits
$g_{1,2}/\Omega \ll \alpha \ll 1$.

In an experiment, the oscillator frequency and the coupling strength are
finite, though. Consequently, the actual phase $\varphi_\mathrm{exp}
(t)$, which we extract numerically and which can be
measured by lock-in amplification, generally differs from the
predicted phase $\varphi (t)$. Thus, the mutual agreement of both
phases needs to be tested quantitatively for realistic scenarios. To
this end, we employ the  measurement fidelity $F$ with the scaled
overlap defined as
\begin{equation}
 F = ( \varphi , \varphi_\mathrm{exp} ) 
 \equiv \left[\int \rmd t\, \varphi^2 (t)\int \rmd t\,
   \varphi_\mathrm{exp}^2 (t)\right]^{-1/2} \Big| \int \rmd t\,
 \varphi (t) \varphi_\mathrm{exp} (t)\Big| \; .
 \label{eq:fidelity}
 \end{equation}
The ideal value of $F=1$ is assumed if $\varphi (t) \propto
\varphi_\mathrm{exp} (t) $.
\begin{figure}[tb!]
  \centering
  \includegraphics[scale=1.]{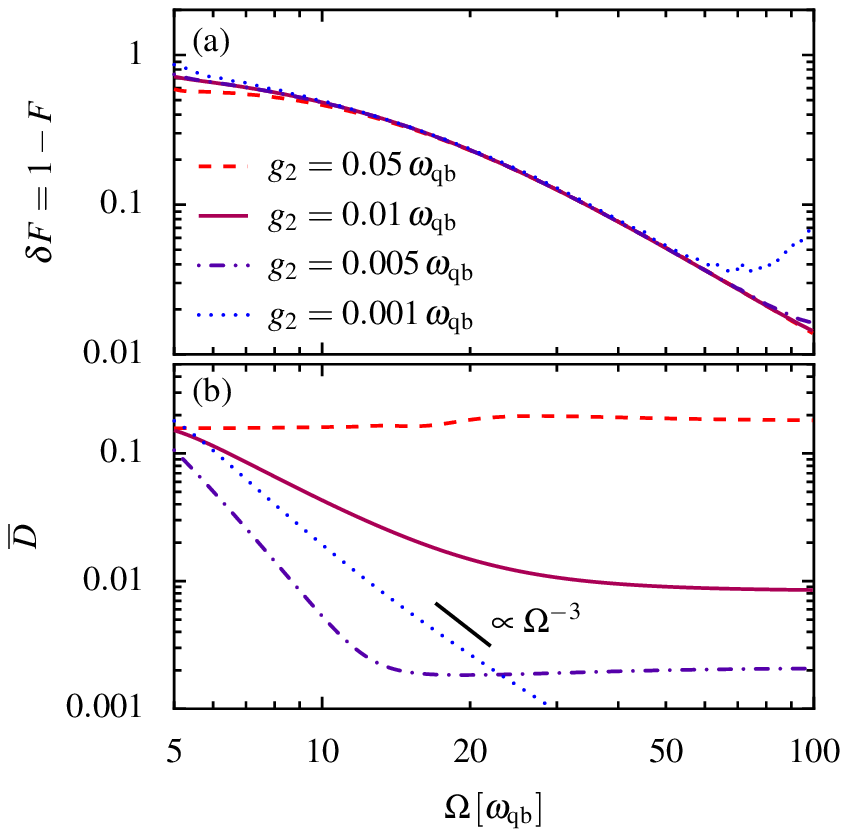}
\caption{(a) Fidelity defect $\delta
  F\,{=}\,1\,{-}\,F$ for the phases $\varphi (t)$ and
  $\varphi_\mathrm{exp} (t)$ and (b) time-averaged trace distance
  $\bar D$ between the density operators of a qubit with finite
  coupling to the oscillator and a reference qubit without
  oscillator. Both quantities are depicted for various coupling
  strengths $g_2$ in dependence of the oscillator frequency
  $\Omega$. All other parameters are as in
  figure~\ref{fig:qubit-osc-phase-spectrum}.}
\label{fig:qubit-osc-fid}
\end{figure}

In figure~\ref{fig:qubit-osc-fid}(a) we depict the fidelity defect
$\delta F = 1-F$ between $\varphi_{\mathrm{exp}}(t)$ and $\varphi (t)$
as a function of the oscillator frequency $\Omega=\Omega_\mathrm{ac}$
for different quadratic coupling coefficients $g_2$. As expected, the
overall fidelity is rather insufficient for small oscillator frequency
$\Omega < 10 \omega_\mathrm{qb}$, for which the adiabatic
approximation of section \ref{sec:disp} is not valid and, moreover, if
the oscillator bandwidth is too small to resolve the qubit dynamics,
i.e., if $\omega_\mathrm{qb} < \alpha \Omega$. Along with increasing
$\Omega$, the fidelity defect $\delta F$ drops to values of
$0.05$--$0.5$, independently of the parameter $g_2$. Taking into
account that the fidelity is arbitrarily lowered by the constant delay
between $\varphi (t)$ and $\varphi_\mathrm{exp} (t)$ visible in
figure~\ref{fig:qubit-osc-phase-spectrum}, this still corroborates
that $\Omega = 10 \omega_\mathrm{qb}$ is a good choice. In the
limit of large oscillator frequencies, we again observe an increase of
the fidelity defect, which occurs the sooner the smaller $g_2$. This
latter effect, which is only visible for the smallest value of $g_2$
in figure~\ref{fig:qubit-osc-fid}(a), is directly explained by a
reduced maximum angular visibility of the phase $\varphi (t) \propto
g_2/\Omega$. Thus, figure~\ref{fig:qubit-osc-fid}(a) provides a
pertinent indication for the validity frame of our central
relation~\eref{eq:dr-osc-phase}. 

Moreover, it is necessary to take into account the backaction upon
the qubit that stems from the non-linear qubit-oscillator interaction.
An appropriate measure for how much the qubit dynamics is perturbed by
the oscillator is given by the time-average $\bar D$ of the trace
distance $D(t) = \frac{1}{2}{\rm Tr}|\rho_\mathrm{qb}(t) -
\rho_\mathrm{qb,0}(t)|$ between the qubit dynamics with and without
the coupling to the driven oscillator. To be specific, we compare the
qubit state $\rho_\mathrm{qb} (t)$ evolving under the full system-bath
Hamiltonian~\eref{eq:sb} to an unperturbed reference state
$\rho_\mathrm{qb,0} (t)$ that evolves unitarily under the bare qubit
Hamiltonian $ \mathcal{H}_\mathrm{qb} = (\hbar \omega_\mathrm{qb} /2)
\sigma_z $. Thus, the trace distance essentially quantifies the
invasiveness of the measurement based upon the second-order
qubit-oscillator interaction. In the absence of perturbations to the
qubit, $\bar D$ vanishes by definition, while $\bar D=1$ if the
density operator of the measured qubit is completely unrelated to that
of the reference.

Figure~\ref{fig:qubit-osc-fid}(b) shows that the predicted phase
$\varphi(t)$ faithfully describes the unperturbed qubit dynamics as
long as the coefficient $g_2$ stays sufficiently small.  A reliable
operating range appears to be $g_2/\omega_\mathrm{qb} \lesssim 0.01$.
For $\Omega = 10\, \omega_\mathrm{qb} $, this is consistent with our
above reasoning regarding the fidelity. For even weaker quadratic
interactions, we first find $\bar D \propto \Omega^{-3}$, which
implies that the dispersive first-order coupling in terms of $g_1$
governs the qubit-oscillator interaction when $\Omega$ is small. This
cubic dependence is due to relation~\eref{eq:dr-osc-phase} and to the
fact that the first-order perturbation acting on the qubit has an
inverse quadratic dependence on the detuning $\Delta \propto \Omega$.
Beyond a critical detuning, which individually depends on $g_2$, the
quadratic interaction prevails again, as is reflected by the
saturation of $\bar D$ with increasing oscillator frequency $\Omega$.
For $g_2/\omega_\mathrm{qb} \gtrsim 0.05$, the effect of
the linear coupling withers. Thus, at $g_2 = 0.01 \omega_\mathrm{qb}$,
the value of $g_1$ is rather irrelevant for our measurement protocol.

\subsection{Signal-to-noise ratio}
\label{sec:sn}

Generally, a signal can be resolved only if its spectral density
exceeds the level of background noise at the measurement frequency.
In the present scheme, the desired information is contained in the
sidepeaks of the spectrum at $\Omega\pm\omega_\mathrm{qb}$; see
figure~\ref{fig:qubit-osc-phase-spectrum}(b).  For the corresponding
phase-modulated oscillation $A\cos(\Omega t -\varphi_\mathrm{max}
\sin(\omega_\mathrm{qb}t))$, these sidepeaks correspond in the time domain
to the oscillation $A\varphi_\mathrm{max}\sin(\omega_\mathrm{qb}t)$.
If the signal is integrated for a time $t$, its spectral weight
becomes $(A\varphi_\mathrm{max})^2 t$.  The phase amplitude
$\varphi_\mathrm{max}$ is given by equation~\eref{eq:phaseresolution},
but for the present purpose, it is sufficient to consider the
dominating contribution which is the one proportional to $g_2$.  Thus,
here $\varphi_\mathrm{max} = 4g_2/\alpha\Omega$, while we restrict
ourselves to the case $\theta = \pi/2$.

Since the measured signal corresponds to the state of a highly excited
environmental mode, the relevant noise level is determined by the
fluctuations of the effective bath coordinate $\xi$.  If the
temperature is sufficiently low, such that thermal excitations do not
play any role, its spectral density equals the bath spectral density:
$S_{\xi\xi}(\omega) = J(\omega) = \alpha\omega$.  Thus, the signal is at
least as big as the noise background if $J(\Omega) \leq
(A\varphi_\mathrm{max})^2 t$.  In other words, the time $t$ during
which the output is recorded must fulfil
\begin{equation}
\label{tmin}
t \geq
\frac{4 J(\Omega)}{(A\varphi_\mathrm{max})^2}
= \frac{(\alpha\Omega)^3}{(2 g_2 A)^2}
\equiv t_\mathrm{meas} \;.
\end{equation}
For the parameters used in figure~\ref{fig:qubit-osc-phase-spectrum},
$t_\mathrm{meas} \approx 4 \cdot 10^3/\omega_\mathrm{qb}$.

Since the measurement is performed via a coupling to external degrees of
freedom, the qubit experiences unavoidable decoherence, which means
that coherent qubit oscillations fade away with a decoherence rate
$\gamma_\phi$.  The time during which meaningful information can be
obtained is therefore limited by the inequality $t\leq 1/\gamma_\phi$.
This condition together with condition \eref{tmin} can be fulfilled
only if $\gamma_\phi t_\mathrm{meas} \geq 1$.

The decoherence rate can be estimated upon noticing that our qubit
Hamiltonian represents a generalised spin-boson model
\cite{Leggett1987a, Hanggi1990a, Weiss1999a} with the bath coupling
$\frac{1}{2} \sigma_x \eta$, where $\eta = 2g_2 Q^2$ is the effective
bath coordinate.  For weak dissipation, $\gamma_\phi$ is given by the
auto-correlation function of the latter evaluated at the qubit
splitting, i.e., $\gamma_\phi = C_{\eta\eta}(\omega_\mathrm{qb})$
\cite{Governale2001a}.  This still holds true in the presence of ac
driving provided that the driving-induced renormalization of the qubit
splitting is negligible \cite{Fonseca2004a}.  We separate the qubit
coordinate into the responses to the driving and to the incoming
fluctuations, $Q = \langle Q\rangle +\delta Q$; cf.\
equation~\eref{eq:in-det}. Then we proceed along the lines of
Ref.~\cite{Clerk2010a}: The relevant terms are those of second order
in $\delta Q$, such that $C_{\eta\eta}(t) \approx
4g_2^2(A/\alpha\Omega)^2 \cos(\Omega t) \langle\delta Q(t) \delta
Q(0)\rangle$.  By Fourier transformation, we obtain to lowest order in
$\omega_\mathrm{qb}/\Omega$ the decoherence rate
\begin{equation}
  \label{eq:dephasingrate}
\gamma_\phi = \frac{(2 g_2 A)^2}{(\alpha\Omega)^3} ,
\end{equation}
which is the inverse of the required measurement time
$t_\mathrm{meas}$.
A comparison with the numerically computed decay of the qubit
coherence (not shown) confirms this value.
The obtained relation $\gamma_\phi t_\mathrm{meas} = 1$ marks the
quantum limit of a measurement~\cite{Clerk2010a} and allows one to
marginally fulfil both conditions on the measurement time $t$.

\subsection{A possible experimental implementation}
\label{sec:experimental}

Specific parameters can be obtained for the setup of reference
\cite{Bertet2005b} for which the qubit-oscillator coupling parameters
are determined by the flux bias current $I_\mathrm{b}$ and read
\begin{eqnarray}
  \label{eq:coefficients}
  g_1 & = - \frac{M I_\mathrm{p}}{4 \hbar I_\mathrm{C}} \frac{\sin
    (\varphi/2)}{\cos^2 (\varphi/2)} \sqrt{\frac{\hbar \Omega}{2 L_\mathrm{J}}}
  I_\mathrm{b} \; ,\\ 
  g_2 & = - \frac{M I_\mathrm{p}}{16 L_\mathrm{J} I_\mathrm{C}} \frac{\sin
    (\varphi/2)}{\cos^2 (\varphi/2)} \Omega \;.
\end{eqnarray}
Here, $M$ denotes the mutual SQUID-qubit inductance, $I_\mathrm{p}$
is the qubit persistent-current, $I_\mathrm{C}$ the critical
current of the SQUID Josephson junctions, and $L_\mathrm{J} =
\phi_0[4\pi I_\mathrm{C} \cos(\pi\phi_\mathrm{SQ}/\phi_0)]^{-1}$ the
SQUID Josephson inductance. The flux  
$\phi_\mathrm{SQ}$ that penetrates the SQUID loop corresponds to the
phase $\varphi=2\pi \phi_\mathrm{SQ}/\phi_0$, with $\phi_0=h/2e$ being
the flux quantum. For small bias currents $I_\mathrm{b}$ while
neglecting the inductance of the wire that leads to the shunting
capacitance $C$, the oscillator frequency is approximately given by
the SQUID plasma frequency $\Omega = |L_\mathrm{J}C|^{-1/2}$.

For a typical qubit transition frequency of $\omega_\mathrm{qb}/2\pi =
5\mathrm{GHz}$, resolving the qubit dynamics requires an
oscillator frequency of $\Omega/2\pi = 10 \omega_\mathrm{qb}/2\pi =
50\mathrm{GHz}$. The necessary lock-in measurements at a carrier
frequency $\Omega/2\pi = 50\mathrm{GHz}$ are particularly challenging
at low temperatures, but feasible \cite{Lee2008a}. They require rather
expensive amplifying technology such as cryogenic amplifiers developed
by Low Noise Factory (Sweden). Recently developed suitable devices can
be operated at up to $36\mathrm{GHz}$ and possess reasonably low noise
temperatures.

Alternatively, a Josephson parametric amplifier
\cite{Ojanen2007b,Yamamoto2008a} enables the detection of oscillator
frequencies as high as $\Omega/2\pi = 20$--$25\mathrm{GHz}$.
Thus, as a compromise, we restrict ourselves to an oscillator
frequency of $\Omega/2\pi = 24\mathrm{GHz}$, which is suitable to
detect the dynamics of a qubit with $\omega_\mathrm{qb}/2 \pi =
2.5\mathrm{GHz}$, a value still large enough to avoid thermal
excitations at working temperatures of $20\mathrm{mK}$. The above
value for $\Omega$ can be realised using the parameters $I_\mathrm{C}
= 4.25 \mu\mathrm{A}$, $\varphi=2.3\pi$, and $C=1\mathrm{pF}$, which
are similar to those of references~\cite{Bertet2005a,Bertet2005b}.
Along with $M=17.5\mathrm{pH}$, $I_\mathrm{p} = 300\mathrm{nA}$ and
$I_\mathrm{b} = 0.4 \mu\mathrm{A}$, we obtain the qubit-oscillator
coupling coefficients $g_1/2\pi = 45 \mathrm{MHz}$ and $g_2 = 24
\mathrm{MHz}$. Thus, the relevant dimensionless coupling assumes the
value $g_2/\omega_\mathrm{qb} \simeq 0.01$ used in our numerical
studies.  An adequate oscillator half-bandwidth is $\Delta \Omega =
2.9\mathrm{GHz}$, which implies a low external quality factor of
$Q\simeq 4$--$5$.

Our two-state model for the qubit does not consider possible
excitations to non-qubit states caused by the coupling to the
oscillator.  Nevertheless our modelling is appropriate, because such
leakage has far less relevance than for a Cooper-pair box, owing to
the fact that the higher states couple only weakly to the
SQUID~\cite{Chirolli2009a}. Apart from this, it is possible to design
or tune the oscillator such that its frequency is far from any qubit
resonance.  The required oscillator frequency of the order
$10\,\mathrm{GHz}$ is still significantly smaller than the gap energy
of aluminium, such that quasi particle excitation should not play a
major role.  This issue is even less critical for niobium.

\section{Conclusions}
\label{sec:conclusion}

We have generalised dispersive qubit readout to
time-resolved observation of the qubit dynamics.  Concerning the
setup, the main difference to dispersive readout is that in the
present proposal, the oscillator frequency needs to exceed the qubit 
splitting by roughly one order of magnitude,
and the oscillator bandwidth should be at least twice the
  qubit frequency.  Also here, the
oscillator frequency becomes dynamically red or blue detuned,
depending on the state of the qubit. When driving the SQUID
oscillator at its bare frequency $\Omega$, this detuning turns into a
phase shift visible in the reflected signal via lock-in techniques.
For such qubit measurement using the oscillator phase, the oscillator
frequency represents the sampling rate, which explains the need for
high frequencies.

The constituting measurement relation has been derived from the
input-output formalism under time-scale separation of the bare qubit
dynamics from the oscillator. A numerical solution of the
Bloch-Redfield master equation for the full qubit-oscillator dynamics
allowed us to compute the phase of reflected signal also directly.
Its good agreement with the phase predicted by our measurement
relation confirms the validity of the latter even when the oscillator
frequency is just moderately large. Thus, there is no need for 
driving the qubit with extremely high frequencies, which would be
quite challenging in an experiment. The found agreement is
also reflected by the measurement fidelity, which already for moderate
frequencies is rather good. Furthermore, the numerical analysis has
demonstrated that the external ac-driving does not significantly
modify the qubit dynamics, which means that the backaction of the
measurement process is weak.  However, it must be emphasised that the
whole scheme relies on the coupling of the qubit via the oscillator to
a dissipative environment, which causes qubit decoherence already when
the external driving is not active. In the limit of far
qubit-oscillator detuning, this qubit decoherence gets drastically
reduced though.

Evaluating the measurement relation for parameters of recent
experiments with flux qubits predicts phase shifts up to $2^\circ$,
which can be measured. Moreover, it reveals that the signal
mainly stems from the coupling of the qubit to the square of the
oscillator coordinate.  The linear coupling to the coordinate, by
contrast, leads to a rather small phase shift.  This means that the
measured quantity is essentially the qubit's flux degree of freedom.
Likewise, the linear coupling of a superconducting charge qubit to a
 waveguide resonator is also too weak. Since for this
system the non-linear coupling practically vanishes, the measured
signal remains tiny.  In conclusion, with present technologies, our
measurement protocol should be feasible best with flux qubits coupled
to SQUIDs that possess a significant non-linear Josephson inductance.
All in all, our proposal may initiate further
progress on the way towards single-shot experiments that demonstrate
quantum coherence in solid-state devices.

\section{Acknowledgements}

We thank Klaus M{\o}lmer, Alexander Baust, Frank Deppe,
and Max H\"aberlein for fruitful discussions.
This work has been supported by the German Research Foundation (DFG)
through the Collaborative Research Centre SFB 631 and through the
German Excellence Initiative via the ``Nanosystems Initiative Munich
(NIM)''.  We acknowledge support by the Spanish Ministry of Science
and Innovation (MICINN) through grants no.\ MAT2008/02626 (SK),
no.\ FIS2008-01240 and no.\ FIS2009-13364-C02-0 (DZ).

\appendix

\section{System-bath Hamiltonian in the dispersive coupling limit}
\label{app:dispersive}

In the limit of large oscillator-qubit detuning, the coupling
coefficients automatically fulfil the conditions
\begin{equation}
  g_1,\,g_2  \ll\Delta\,,
  \quad\Delta=\Omega - \omega_\mathrm{qb} \; ,
   \label{eq:drivenosc-dispersive}
\end{equation}
which mark the dispersive coupling regime. Following
references~\cite{Klimov2000a,Klimov2003a}, the effective
Hamiltonian $\mathcal{H}_\mathrm{0,disp} = \mathcal{U}^\dagger
\mathcal{H}_0 \mathcal{U}$ is then obtained from the full system
Hamiltonian~\eref{eq:H_qc} by the unitary transformation
\begin{equation}
\label{eq:drivenosc-Schrieffer}
  \mathcal{U} = \exp\big(
  \lambda_\Delta \mathcal{D}
               + \lambda_\Sigma \mathcal{S}
               + \lambda_\Omega\mathcal{W}\big) \; ,
\end{equation}
where
\begin{eqnarray}
\mathcal{D} &= \sigma^- a^\dagger - \sigma^+ a \label{eq:drivenosc-X-} \;, \\
\mathcal{S} &= \sigma^- a - \sigma^+ a^\dagger \label{eq:drivenosc-Y-} \;, \\
\mathcal{W} &= \sigma_z (a - a^\dagger) \label{eq:drivenosc-Z-} \; .
\end{eqnarray}
Defining $\Sigma = \omega_\mathrm{qb} + \Omega$, the necessarily small
and dimensionless dispersive parameters
\begin{eqnarray}
  \lambda_\Delta &= - \frac{g_1 \sin \theta}{\Delta} \,,
\label{eq:drivenosc-lambdaX-}\\
 \lambda_\Sigma &= \hphantom{-} \frac{g_1 \sin \theta}{\Sigma} \,,
\label{eq:drivenosc-lambdaY-} \\
\lambda_\Omega &= - \frac{g_1 \cos \theta}{\Omega} 
 \label{eq:drivenosc-lambdaZ-}
\end{eqnarray}
emerge.  Expanding the transformed Hamiltonian in powers of
$\lambda_{\Delta,\Sigma,\Omega}$, we obtain to second dispersive order
the effective Hamiltonian
\begin{equation}
  \label{eq:drivenosc-Heff}
\eqalign{
    \bar  {\mathcal H}_0
= {} &
    \hbar\Omega\Big(a^\dagger a+ 
    \frac{1}{2}\Big) + \frac{\hbar\omega_\mathrm{qb}}{2}\sigma_z
    +\frac{\hbar}{2}\big(\Delta\lambda_\Delta^2 -
    \Sigma\lambda_\Sigma^2\big)
    \sigma_z \big (a + a^\dagger \big)^2
\\
    & {}+\frac{\hbar\Omega}{2} \lambda_\Omega \left(\lambda_\Delta +
      \lambda_\Sigma\right) \sigma_x \big (a + a^\dagger\big)^2 -
    \frac{\rmi\omega_\mathrm{qb}}{2} \lambda_\Omega \left(\lambda_\Delta +
      \lambda_\Sigma\right) \sigma_y  \Big(a^2 -  (a^\dagger)^2 \Big)
\\
    & + g_2 (\cos \theta \sigma_z - \sin \theta \sigma_x) \big (a +
    a^\dagger\big)^2 \, .
}
\end{equation}
The third and fourth term of this Hamiltonian constitute corrections
to the curvature of the oscillator potential, i.e., the prefactor of
$(a + a^\dagger)^2$.  They stem from the linear qubit-oscillator
interaction and, thus, enter only in second dispersive order. Since we
consider a high-frequency oscillator, the detuning is always
positive, $\Delta >0$, such that the dispersive parameters
$\lambda_\Delta$ and $\lambda_\Sigma$ are of opposite sign.  Thus, in
the case of far dispersive detuning $\Omega \gg \omega_\mathrm{qb}$,
these terms become rather small. In spite of this, we keep them up for
later purpose. On the contrary, we can safely neglect the fifth term
which is not of the shape $(a + a^\dagger)^2$ and whose coefficient is
small as compared to the other terms.

The last term of equation~\eref{eq:drivenosc-Heff}, stemming from the
second-order interaction between the qubit and the oscillator, plays a
particular role. Since it already is of second order in the oscillator
coordinate $a+a^\dagger$ and its coefficient $g_2$ is correspondingly
small, $g_2 \ll g_1$, it is not affected by the
transform~\eref{eq:drivenosc-Schrieffer}. As a consequence, this term
remains independent of the qubit-oscillator detuning $\Delta$, for
which reason its contribution to the $(a + a^\dagger)^2$-terms is
finite.

For further convenience, we introduce transformed creation and
annihilation operators that describe the oscillator-qubit system  in
the adiabatic limit $\Omega \gg \omega$,
\begin{equation}
  \label{eq:abar}
\eqalign{
    {\bar a} &= \frac{1}{2} \sqrt{\frac{\bar \Omega}{\Omega}} (a +
    a^\dagger) + \frac{1}{2} \sqrt{\frac{ \Omega}{\bar \Omega}}
    (a - a^\dagger) ,
}
\end{equation}
and $\bar a^\dagger$ accordingly, such that
$[\bar a,\bar a^\dagger] = 1$. The effective oscillator frequency
\begin{equation}
  \label{eq:Omegabar}
  \bar \Omega = \Omega  \sqrt{1 + \frac{4 \bar
      \omega}{\Omega}}
\end{equation}
accounts for all quadratic corrections to the oscillator potential in
the effective Hamiltonian~\eref{eq:drivenosc-Heff} in terms of 
the effective operator-valued coupling frequency
\begin{equation}
  \label{eq:omegabar}
    \bar \omega  = \frac{1}{2}(\Delta\lambda_\Delta^2 -
    \Sigma\lambda_\Sigma^2) \sigma_z 
    +\frac{1}{2}\Omega \lambda_\Omega ( \lambda_\Delta +
      \lambda_\Sigma)  \sigma_x   
     + g_2 (\cos \theta  \sigma_z  - \sin\theta \sigma_x ) \;.
\end{equation}
Thus, the effective Hamiltonian can be rewritten as 
\begin{equation}
  \label{eq:drivenosc-Heff1}
   \bar  {\mathcal H}_0=  \hbar\bar
   \Omega \Big( \bar a^\dagger \bar a + \frac{1}{2}\Big) 
    + \frac{1}{2}\hbar\omega_\mathrm{qb}\sigma_z \; .
\end{equation}
Put differently, the qubit-oscillator coupling has been shifted to the
effective operator-valued oscillator frequency $\bar\Omega$, which
depends on the qubit state.

In order to move fully to the dispersive picture, we also have to
transform the system operator $Q = a + a^\dagger$ by which the
oscillator couples to the environment.  
Transformation with the operator~\eref{eq:drivenosc-Schrieffer} yields
in first dispersive order the position operator
\begin{equation}
  \label{eq:drivenosc-Qeff}
    \bar Q
=  \mathcal{U}^\dagger Q \mathcal{U}
=  (\bar a + \bar a^\dagger) -(\lambda_\Delta -
    \lambda_\Sigma)\sigma_x + 2 \lambda_\Omega \sigma_z
\; ,
\end{equation}
where we have assumed $\Omega \approx \bar\Omega$.
The full system-bath Hamiltonian in the dispersive picture finally
reads as
\begin{equation}
  \label{eq:drivenosc-sbeff}
    \bar  {\mathcal H} 
   =   \bar  {\mathcal H}_0  +
  \hbar  \bar Q  \sum_n c_n 
   (b_n^{\vphantom{\dagger}} + b_n^{\dagger}) + \sum_n \hbar \omega_n
   \Big( b_n^{\dagger}b_n^{\vphantom{\dagger}} + \frac{1}{2} \Big)
   \; .
\end{equation}

\section{Input-output formalism}
\label{app:input-output}

In order to compute the response of the oscillator to the external
driving, we employ the input-output formalism \cite{Gardiner1985b},
which is most conveniently obtained from the quantum Langevin equation
of the central system
\cite{Magalinskii1959a,Benguria1981a,Ford1987a,Hanggi2005b}. We
derive it from the system-bath Hamiltonian \eref{eq:sb} via the Heisenberg
equation of motion for the bath oscillator coordinates $q_n =
b_n^{\vphantom{\dagger}} + b_n^\dagger$,
\begin{equation}
  \label{eq:drivenosc-bathmodes-eq}
  \ddot q_n +\omega_n^2 q^{\vphantom{\dagger}}_n = 2 c_n \omega_n Q  \; .
\end{equation}
Here, the oscillator coordinate $Q = a + a^\dagger$ enters as
inhomogeneity. The formal solution of
equation~\eref{eq:drivenosc-bathmodes-eq} for initial time $t_0$ is
\begin{equation}
  \label{eq:drivenosc-bathmodes-sol}
\eqalign{
    q_n (t) = {} & q_n (t_0) \cos \{\omega_n (t-t_0)\} + \frac{p_n
      (t_0)}{\omega_n} \sin \{\omega_n (t-t_0)\} \\
    & + 2 c_n \int_{t_0}^t \rmd t' \, \sin\{\omega_n (t-t')\} Q (t') \; ,
}
\end{equation}
with $p_n (t_0) = \dot q_n (t_0)$.  Inserting this solution into the
Heisenberg equation of motion for $Q$ yields
\begin{equation}\label{eq:drivenosc-sysop-eq}
\eqalign{
    \ddot{Q} 
={} & -  \Omega^2 Q - 4  \Omega \sum_n c_n^2 \int_{t_0}^t \rmd t'
    \sin\{\omega_n(t-t')\}  Q(t')
\\
    & - 2 \Omega \sum_n c_n \left( q_n(t_0) \cos\{\omega_n (t-t_0)\}
   + \frac{p_n (t_0)}{\omega_n} \sin\{\omega_n (t-t_0)\}
    \right) .
}
\end{equation}
For the sake of a compact notation, we define the operator for the
incoming fluctuations,
\begin{equation}
  \label{eq:drivenosc-xi-in}
  \xi_\mathrm{in}^{\rm qm} (t)
=  \sum_n c_n  \left( q_n(t_0) \cos\{\omega_n (t-t_0)\}
   + \frac{p_n (t_0)}{\omega_n} \sin \{\omega_n (t-t_0)\} \right) \, ,
\end{equation}
which only depends on the environmental operators at initial time and,
thus, is independent of the central quantum system.

We replace the sum $\sum_n|c_n|^2$ by an integral over the spectral
density $J(\omega)/\pi$, which for the ohmic $J(\omega)=\alpha\omega$
becomes the derivative of the delta function $\delta(t-t')$, such that
the time integral can be evaluated.  In doing so, we arrive at the
quantum Langevin equation
\begin{equation}\label{eq:drivenosc-langevin-in1}
  \ddot { Q} + 2 \alpha  \Omega \dot{Q} +  \Omega^2  Q
= - 2 \Omega \xi_\mathrm{in}^{\rm qm}(t) ,
\end{equation}
where we have discarded an initial slip term and a constant potential
renormalisation which both are not relevant in the present context and
beyond transient behaviour. Notice that dissipation enters via a
friction term, while the incoming fluctuations act as stochastic
driving force.

The quantum Langevin equation~\eref{eq:drivenosc-langevin-in1} can
also be expressed in terms of the outgoing fluctuations by solving the
equations of motion~\eref{eq:drivenosc-bathmodes-eq} for $q_n$ with
boundary condition at a later time time $t_1>t$, i.e., by
backward propagation. Then one obtains
\begin{equation}
  \label{eq:drivenosc-bathmodes-sol-II}
\eqalign{
    q_n (t) 
={} & q_n (t_1) \cos\{\omega_n (t-t_1)\}
    + \frac{p_n (t_1)}{\omega_n} \sin\{\omega_n (t-t_1)\}
\\ &
  + 2 c_n \int_t^{t_1} \rmd t' \, \sin\{\omega_n (t-t')\} Q (t') \; .
}
\end{equation}
The corresponding environment operators define the outgoing
fluctuations
\begin{equation}
  \label{eq:drivenosc-xi-out}
  \xi_\mathrm{out}^{\rm qm} (t) =  \sum_n c_n  
\left( q_n(t_1) \cos\{\omega_n (t-t_1)\}
  + \frac{p_n (t_1)}{\omega_n} \sin\{\omega_n (t-t_1)\} \right) \, .
\end{equation}
In contrast to $\xi_\mathrm{in}^{\rm qm}(t)$, this noise operator
depends on the time evolution of the system at earlier times $t <
t_1$.  The resulting Langevin equation for the oscillator
coordinate $Q$,
\begin{equation}\label{eq:drivenosc-langevin-out}
  \ddot {Q} - 2 \alpha  \Omega \dot {Q} 
  +  \Omega^2  Q = - 2  \Omega
  \xi_\mathrm{out}^{\rm qm} (t)
\end{equation}
is characterised by negative damping and the outgoing noise.
The difference of both Langevin equations links the noise terms via
twice the dissipative term by means of the input-output relation,
which in the stationary limit $t_0\to-\infty$ and $t_1\to\infty$ reads
\cite{Gardiner1985b}
\begin{equation}\label{eq:in-out-a0}
  \xi_{\mathrm{out}}^{\rm qm} (t) - \xi_{\mathrm{in}}^{\rm qm} (t)
= 2 \alpha  \dot{Q} \, .
\end{equation}
Even though we have written this relation for a harmonic oscillator,
the derivation does not rely on particular properties of this
system.  Thus, equation \eref{eq:in-out-a0} is valid as well for
non-linear quantum systems coupled to an environment.

If a bath mode is coherently excited by an external driving field, the
incoming fluctuations are augmented by a deterministic contribution,
$\xi_\mathrm{in}^\mathrm{qm} \to \xi_\mathrm{in}^\mathrm{qm} +
x_\mathrm{drive}(t)$.  Then the input-output relation allows one to
compute both the averaged outgoing signal as well as its fluctuations
and noise spectra.

\section{Bloch-Redfield master equation}
\label{app:QME}

The numerical data presented in section~\ref{sec:numerics} have been
computed with a quantum master equation of the Bloch-Redfield
type~\cite{Blum1996a},
\begin{equation}
  \label{eq:blochredfield}
    \dot \rho_0 (t) =
    -\frac{\rmi}{\hbar}\big[\mathcal H_0,\rho_0 (t) \big]
  -  \big[ Q, \big[\hat Q, \rho_0  (t)\big]  \big]
  + \rmi \alpha
  \big[ Q, \big\{ \dot Q, \rho_0  (t) \big\}  \big]
 \; ,
 \end{equation}
where
\begin{equation}\label{eq:2res-Qhat}
  \hat Q = \frac{\alpha}{\pi} \int_0^\infty \rmd\tau \int_0^\infty
  \rmd\omega\, \omega \coth \Big(\frac{\hbar \omega}{2 k_\mathrm{B} T}
  \Big)  \cos(\omega\tau) \tilde Q (-\tau) \; .
\end{equation}
It describes the time-evolution of the reduced density operator
$\rho_0 (t)$ of the qubit plus the oscillator.  The dissipative terms
have been derived under the assumption that the bath couples
weakly to a system operator $Q$ with a vanishing equilibrium
expectation value.  The environment is in a thermal state at
temperature $T$, and the system-bath interaction possesses the ohmic
spectral density $J (\omega)=\alpha\omega$ with the dimensionless
damping strength $\alpha$. Furthermore, $\tilde X (t) = \mathcal
U_0^\dagger (t,t_0) X \mathcal U_0 (t,t_0)$ refers to the time
evolution of the system operator $X$ in an interaction picture
described by the propagator $\mathcal U_0 (t,t_0)= \exp\{\rmi \mathcal
H_0 (t-t_0)/\hbar\}$, and $\dot Q$ is a shorthand notation for the
Heisenberg time derivative $\rmi [\mathcal H_0,Q]/\hbar$ of the
system-bath coupling operator $Q$.

\section*{References}

\bibliographystyle{iopart-num}

\begin{thebibliography}{10}
\expandafter\ifx\csname url\endcsname\relax
  \def\url#1{{\tt #1}}\fi
\expandafter\ifx\csname urlprefix\endcsname\relax\def\urlprefix{URL }\fi
\providecommand{\eprint}[2][]{\url{#2}}

\bibitem{Zurek2003a}
Zurek W~H 2003 {\em Rev. Mod. Phys.\/} {\bf 75} 715

\bibitem{Nakamura1999a}
Nakamura Y, Pashkin Y~A and Tsai J~S 1999 {\em Nature (London)\/} {\bf 398} 786

\bibitem{Leek2007a}
Leek P~J, Fink J~M, Blais A, Bianchetti R, G{\"o}ppl M, Gambetta J~M, Schuster
  D~I, Frunzio L, Schoelkopf R~J and Wallraff A 2007 {\em Science\/} {\bf 318}
  1889

\bibitem{Ansmann2009a}
Ansmann M, Wang H, Bialczak R~C, Hofheinz M, Lucero E, Neeley M, O'Connell A~D,
  Sank D, Weides M, Wenner J, Cleland A~N and Martinis J~M 2009 {\em Nature
  (London)\/} {\bf 461} 504

\bibitem{Sillanpaa2005a}
Sillanp\"a\"a M~A, Lehtinen T, Paila A, Makhlin Y, Roschier L and Hakonen P~J
  2005 {\em Phys. Rev. Lett.\/} {\bf 95} 206806

\bibitem{Lupascu2006a}
Lupascu A, Driessen E~F~C, Roschier L, Harmans C~J~P~M and Mooij J~E 2006 {\em
  Phys. Rev. Lett.\/} {\bf 96} 127003

\bibitem{Lupascu2007a}
S A~L, Saito, Picot T, de~Groot P~C, Harmans C~J~P~M and Mooij J~E 2007 {\em
  Nature Phys.\/} {\bf 3} 119

\bibitem{Schuster2007a}
Schuster D~I, Houck A~A, Schreier J~A, Wallraff A, Gambetta J~M, Blais A,
  Frunzio L, Majer J, Johnson B, Devoret M~H, Girvin S~M and Schoelkopf R~J
  2007 {\em Nature (London)\/} {\bf 445} 515

\bibitem{Ashhab2009a}
Ashhab S, You J~Q and Nori F 2009 {\em Phys. Rev. A\/} {\bf 79} 032317

\bibitem{Ashhab2009b}
Ashhab S, You J~Q and Nori F 2009 {\em New J. Phys.\/} {\bf 11} 083017

\bibitem{Chiorescu2004a}
Chiorescu I, Bertet P, Semba K, Nakamura Y, Harmans C~J~P~M and Mooij J~E 2004
  {\em Nature (London)\/} {\bf 431} 159

\bibitem{Wallraff2004a}
Wallraff A, Schuster D~I, Blais A, Frunzio L, Huang R~S, Majer J, Kumar S,
  Girvin S~M and Schoelkopf R~J 2004 {\em Nature (London)\/} {\bf 431} 162

\bibitem{Grajcar2004a}
Grajcar M, Izmalkov A, Il'ichev E, Wagner T, Oukhanski N, H\"ubner U, May T,
  Zhilyaev I, Hoenig H~E, {Y}a S~Greenberg, Shnyrkov V~I, Born D, Krech W,
  Meyer H~G, {Maassen van den Brink} A and Amin M~H~S 2004 {\em Phys. Rev. B\/}
  {\bf 69} 060501(R)

\bibitem{Johansson2006a}
Johansson G, Tornberg L and Wilson C~M 2006 {\em Phys. Rev. B\/} {\bf 74}
  100504(R)

\bibitem{Filipp2009a}
Filipp S, Maurer P, Leek P~J, Baur M, Bianchetti R, Fink J~M, G{\"o}ppl M,
  Steffen L, Gambetta J~M, Blais A and Wallraff A 2009 {\em Phys. Rev. Lett.\/}
  {\bf 102} 200402

\bibitem{Reuther2009a}
Reuther G~M, Zueco D, H{\"a}nggi P and Kohler S 2009 {\em Phys. Rev. Lett.\/}
  {\bf 102} 033602

\bibitem{Reuther2011a}
Reuther G~M, Zueco D, H\"anggi P and Kohler S 2011 {\em Phys. Rev. B\/} {\bf
  83} 014303

\bibitem{Bertet2005a}
Bertet P, Chiorescu I, Burkard G, Semba K, Harmans C~J~P~M, DiVincenzo D~P and
  Mooij J~E 2005 {\em Phys. Rev. Lett.\/} {\bf 95} 257002

\bibitem{Bertet2005b}
Bertet P, Chiorescu I, Harmans C~J~P~M and Mooij J~E 2005 Dephasing of a
  flux-qubit coupled to a harmonic oscillator arXiv:cond-mat/0507290

\bibitem{Gardiner1985b}
Gardiner C~W and Collett M~J 1985 {\em Phys. Rev. A\/} {\bf 31} 3761

\bibitem{Serban2008a}
Serban I, Plourde B~L~T and Wilhelm F~K 2008 {\em Phys. Rev. B\/} {\bf 78}
  054507

\bibitem{Blais2004a}
Blais A, Huang R~S, Wallraff A, Girvin S~M and Schoelkopf R~J 2004 {\em Phys.
  Rev. A\/} {\bf 69} 062320

\bibitem{Mariantoni2008a}
Mariantoni M, Deppe F, Marx A, Gross R, Wilhelm F~K and Solano E 2008 {\em
  Phys. Rev. B\/} {\bf 78} 104508

\bibitem{Leggett1987a}
Leggett A~J, Chakravarty S, Dorsey A~T, Fisher M~P~A, Garg A and Zwerger W 1987
  {\em Rev. Mod. Phys.\/} {\bf 59} 1

\bibitem{Hanggi1990a}
H\"anggi P, Talkner P and Borkovec M 1990 {\em Rev. Mod. Phys.\/} {\bf 62} 251

\bibitem{Weiss1999a}
Weiss U 1998 {\em Quantum Dissipative Systems\/} 2nd ed (Singapore: World
  Scientific)

\bibitem{Makhlin2001b}
Makhlin Y, Sch\"on G and Shnirman A 2001 {\em Rev. Mod. Phys.\/} {\bf 73}
  357--400

\bibitem{Ingold1992a}
Ingold G~L and {Y}u V~Nazarov 1992 Charge tunneling rates in ultrasmall
  junctions {\em Single Charge Tunneling\/} ({\em NATO ASI Series B\/} vol 294)
  ed Grabert H and Devoret M~H (New York: Plenum) pp 21--107

\bibitem{Yurke1984a}
Yurke B and Denker J~S 1984 {\em Phys. Rev. A\/} {\bf 29} 1419

\bibitem{Devoret1995a}
Devoret M~H 1995 {\em Quantum Fluctuations in Electrical circuits\/}
  (Amsterdam: Elsevier) chap~10 Les Houches, Session LXIII

\bibitem{Makhlin2001a}
Makhlin Y and Mirlin A~D 2001 {\em Phys. Rev. Lett.\/} {\bf 87} 276803

\bibitem{Klimov2000a}
Klimov A~B and Sanchez-Soto L~L 2000 {\em Phys. Rev. A\/} {\bf 61} 063802

\bibitem{Klimov2003a}
Klimov A~B, Sainz I and Chumakov S~M 2003 {\em Phys. Rev. A\/} {\bf 68} 063811

\bibitem{Zueco2009b}
Zueco D, Reuther G~M, Kohler S and H{\"a}nggi P 2009 {\em Phys. Rev. A\/} {\bf
  80} 033846

\bibitem{Gardiner2004a}
Gardiner C~W and Zoller P 2004 {\em Quantum Noise\/} 3rd ed (Berlin and
  Heidelberg: Springer)

\bibitem{Scofield1994a}
Scofield J~H 1994 {\em Am. J. Phys.\/} {\bf 62} 129

\bibitem{Governale2001a}
Governale M, Grifoni M and Sch\"on G 2001 {\em Chem. Phys.\/} {\bf 268}
  273--283

\bibitem{Fonseca2004a}
Fonseca-Romero K~M, Kohler S and H\"anggi P 2004 {\em Chem. Phys.\/} {\bf 296}
  307

\bibitem{Clerk2010a}
Clerk A~A, Devoret M~H, Girvin S~M, Marquardt F and Schoelkopf R~J 2010 {\em
  Rev. Mod. Phys.\/} {\bf 82} 1155

\bibitem{Lee2008a}
Lee J, Liu M and Wang H 2008 {\em IEEE J. Solid-State Circuits\/} {\bf 43} 1414

\bibitem{Ojanen2007b}
Ojanen T and Salo J 2007 {\em Phys. Rev. B\/} {\bf 75} 184508

\bibitem{Yamamoto2008a}
Yamamoto T, Inomata K, Watanabe M, Matsuba K, Miyazaki T, Oliver W~D, Nakamura
  Y and Tsai J~S 2008 {\em Appl. Phys. Lett.\/} {\bf 93} 042510

\bibitem{Chirolli2009a}
Chirolli L and Burkard G 2009 {\em Phys. Rev. B\/} {\bf 80} 184509

\bibitem{Magalinskii1959a}
Magalinskii V~B 1959 {\em Zh. Eksp. Teor. Fiz.\/} {\bf 36} 1942 [Sov. Phys.
  JETP {\bf 9}, 1381 (1959)]

\bibitem{Benguria1981a}
Benguria R and Kac M 1981 {\em Phys. Rev. Lett.\/} {\bf 46} 1

\bibitem{Ford1987a}
Ford G~W and Kac M 1987 {\em J. Stat. Phys.\/} {\bf 46} 803

\bibitem{Hanggi2005b}
H\"anggi P and Ingold G~L 2005 {\em Chaos\/} {\bf 15} 026105

\bibitem{Blum1996a}
Blum K 1996 {\em Density Matrix Theory and Applications\/} 2nd ed (New York:
  Springer)

\end{thebibliography}

\providecommand{\newblock}{}

\end{document}